\documentstyle[prl,aps,epsf]{revtex}
\begin{document}
\draft
\twocolumn[\hsize\textwidth\columnwidth\hsize\csname @twocolumnfalse\endcsname

\title{Topological Invariants in Microscopic Transport on
Rough Landscapes: Morphology, Hierarchical Structure, 
and Horton Analysis of River-like Networks of Vortices}
\author{A.P.~Mehta, C.~Reichhardt, C.J.~Olson 
 and Franco Nori$^*$}
\address{Department of Physics, 
The University of Michigan,
Ann Arbor, Michigan 48109-1120}

\date{\today}
\maketitle
\begin{abstract}
River basins as diverse as the Nile, the Amazon, and the Mississippi 
satisfy certain topological invariants known as Horton's Laws.
Do these macroscopic (up to $10^3$ km) laws extend to the micron scale?
Through realistic simulations, we analyze the 
morphology and hierarchical properties of networks 
of vortex flow in flux-gradient-driven superconductors.
We derive a phase diagram of the different 
network morphologies, including one in which 
Horton's laws of length and stream number are 
obeyed---even though these networks are about 
$10^{9}$ times smaller than geophysical river basins.
\end{abstract}
\vspace{-8pt}
\vspace*{-0.3in}
\vskip2pc]
\narrowtext

{\it Introduction.---\/}
The nature of river 
basins\cite{Rodriguez,riverbooks,Horton,RiversRivers}, 
including their physical structure and evolution, 
has been a problem of major interest to civilized 
societies throughout history.   Horton's laws are 
perhaps one of the most intriguing 
properties of river networks \cite{Rodriguez,riverbooks,Horton,RiversRivers}.
In order to apply them to a network, the individual streams 
composing the network must be identified and labeled with an order 
number, as in the top left corner of Fig.~1(a).
The lowest order streams are the smallest outlying tributaries on the
edges of the network, according to the Strahler ordering scheme.
At each point where two tributary streams join, a new stream begins. 
Whenever two tributaries of the same order meet, the outgoing
stream has an order number one higher than that of the tributaries.
If two tributaries of different orders meet, the outgoing stream
has the same order number as the higher ordered tributary.
Eventually, all streams in the network combine to form the 
highest order (main) stream.
The number of streams of order $w$ is $N_{w}$,
while $L_{w}$ is the average length of streams of order $w$.
Horton's laws state that the bifurcation ratio
$R_{B}$ and the length ratio $R_{L}$, given by
$R_{B} = N_{w} / N_{w+1}$ and 
$R_{L} = L_{w+1}/L_{w}$,
are constant, or independent of $w$. 
These ratios also provide the 
fractal dimension \cite{Rodriguez,riverbooks,Horton}
of the rivers
$D_{F} \approx log{R_{B}}/\log{R_{L}}$.
Geophysical river basins 
\cite{Rodriguez,riverbooks,Horton}
typically have values of 
$R_{B} \approx 4$ and $R_{L} \approx 2$.  
Do these (Horton's) laws apply to microscopic landscapes?
%
Here we present evidence that these macroscopic laws are 
obeyed at the microscopic scale by river-like networks 
of flowing quantized magnetic flux.

{\it Vortex River Basins.---} Near the depinning transition, 
magnetic vortices in type-II
superconductors move in intricate flow patterns that have been
seen both in computer simulations
and in experiments, including finger-like or dendritic shapes
as well as the filamentary flow of vortices in river-like paths 
and networks 
(see, e.g., \cite{vortexrivers,RRiver,RBose} and
references therein).
Despite the ubiquity of the river-like pathways produced by the
vortex motion, very little work has been done towards
characterizing the morphology of these flow patterns.
Moreover, concepts and ideas used for decades to characterize 
geophysical river basins have {\it not\/} been applied 
%
%
to the study of the microscopic flow through tree--shaped 
channel networks. This is surprising since the underlying physics
of vortex and geological rivers offers striking {\it similarities}: 
driven non-equilibrium dissipative systems displaying branched 
(or ramified) transport among 
metastable states on a rough landscape \cite{soc}.  
One is driven by the Lorentz force and the other by gravity.
Like geophysical rivers, vortex flow basins exhibit 
sinuosity (i.e., tortuosity), anabranching, braiding, occasional 
sudden floods, and other features that make them remarkably
similar to geophysical rivers \cite{Rodriguez}. 
Indeed, some satellite photographs of river basins are 
strikingly similar to the channels produced by vortex motion.
However, significant {\it differences\/} also exist, including:
flow direction, 
quantized flux flow versus continuum water flow,
compressible vortex lattice versus incompressible fluid,
negligible inertia with overdamped vortex dynamics versus massive fluid,
non-erosional versus erosional landscape,
peripheral flux sources versus uniform rain, and
correlated long-range versus short-range interactions
(so the rapidly varying vortex--vortex repulsion 
landscape smoothes out the underlying static pinscape).
%
%
This strongly-correlated vortex dynamics generates flux motion 
that can be either continuous-flow type, like water, or 
intermittent stick-slip-type motion---depending on the 
balance of forces.
Also, vortices typically move over relatively flat
landscapes with many divots, as opposed to the 
mountain-range-like very rough landscapes 
of some geophysical rivers.
Moreover, vortex river basins occur inside materials
at approximate scales between 1 to 100 $\mu$m,
much smaller than geophysical river basins
(of up to $10^3$ km)---and also spanning a smaller 
range of length scales.
Thus, given these {\it numerous similarities and differences},
it is very unclear {\it a priori\/} which macroscopic 
results carry over to the microscopic domain.

By conducting realistic simulations of slowly
driven vortices moving over many samples,
we have identified several distinct network phases.  
These vortex basins appear in the initial penetrating 
front of vortices\cite{RRiver,RBose}.
Remarkably, we find that: for a wide range of parameters
{\it networks of vortex channels obey Horton's laws\/} just 
as geophysical river networks do.
This is remarkable, given the many physical differences
between basins of flux quanta and geophysical rivers
and that they move over very different types 
of potential-energy landscapes.
%
Unlike previous work, here we first present a detailed 
list of analogies and differences between river basins 
and networks of vortex channels.  Afterwards, we present 
the first {\it morphological\/} phase diagram for vortex 
motion.  Finally, we analyze the hierarchical structure of 
the vortex channels.

\begin{figure}
\centerline{
\epsfxsize=3.3in
\epsfbox{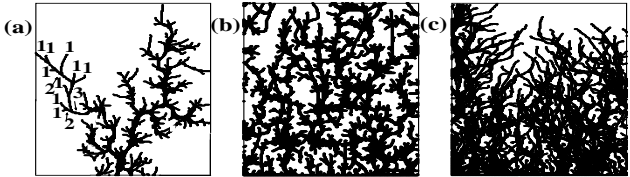}}
\caption{
Snapshots of northbound vortex pathways:
(a) Hortonian
(when the pinning force, $f_p \,$, 
is stronger than the vortex-vortex 
repulsion ${\bf f}^{vv} \,$; i.e., 
at low $B$ and high $f_p$);
(b) braided
(when $f_p$ is comparable to ${\bf f}^{vv} \,$: 
$B \approx 3 B_{\phi} / 2$ [7]); and (c) dense 
(when $f_p  \ll  f^{vv} \,$; 
i.e., for $B > 2 B_{\phi}$; 
or at any field for low $f_p$).
Here, $n_{p}=0.75/\lambda^{2}$.
%
%
The matching field $B_{\phi}$ occurs when the number of 
vortices $N_v$ equals the number of pinning sites $N_p$.  
$B_{\phi}$ is used to quantify the relative strength 
of pinning versus vortex-vortex repulsion [7].
}
\label{fig:rivers}
\end{figure}

{\it Simulation.---\/}
We model a transverse 2D 
slice (in the $x$--$y$ plane) of an infinite zero-field-cooled $T=0$
superconducting slab containing flux-gradient-driven 3D rigid
vortices that are parallel to the sample edge \cite{RRiver,RBose}.  
Vortices are added at the surface 
at periodic time intervals, and enter the superconducting slab 
under the force of their own mutual repulsion  \cite{RRiver,RBose}.  
The slab is $36\lambda \times 36\lambda$ in size, where $\lambda$ is
the penetration depth. 
%
The vortex-vortex repulsive interaction is correctly modeled by a
modified Bessel function, $K_{1}(r/\lambda)$.
The vortices also interact with 972 non-overlapping attractive parabolic 
wells of radius $\xi_{p}=0.3\lambda$.
The density of pins $n_{p}$ is $n_{p}=0.75/\lambda^{2}$.  All pins in
a given sample have the same maximum pinning force $f_{p}$, which ranged
from $f_{p}=0.3f_{0}$ to $f_{p}=6.0f_{0}$ in thirteen different samples.
For each sample type, we considered five realizations of disorder.
Thus, the five points at each pinning force, delineating
the broad crossover boundary between Hortonian and braided 
phases in Fig.~2, refer to these five realizations of disorder.
A sixth point, indicating the average value from the five trials,
is not visible when it overlaps with another point.
%
%
We measure all forces in units of $f_{0}=\Phi_{0}^{2}/8\pi^{2}\lambda^{3}$,
magnetic fields in units of $\Phi_{0}/\lambda^{2}$, 
and lengths in units of the penetration depth $\lambda$.
Here, $\Phi_0$ is the flux quantum.

The overdamped equation of vortex motion is
$ {\bf f}_{i} = {\bf f}_{i}^{vv} + {\bf f}_{i}^{vp}= \eta{\bf v}_{i}$,
where the total force $ {\bf f}_{i}$ on vortex $i$ (due to other
vortices ${\bf f}_{i}^{vv}$, and pinning sites ${\bf f}_{i}^{vp}$)
is given by
${\bf f}_{i} $
$= \sum_{j=1}^{N_{v}}f_{0} \ K_{1}(|{\bf r}_{i} - {\bf r}_{j}|/
\lambda){\bf {\hat r}}_{ij} +$
$\sum_{k=1}^{N_{p}}(f_{p}/\xi_{p})|{\bf r}_{i} - {\bf r}_{k}^{(p)}|$ 
$\ \Theta \left[ \xi_{p} - |{\bf r}_{i} - {\bf r}_{k}^{(p)}| \right]
{\hat{\bf r}}_{ik} \, .$
Here, $\Theta$ is the Heaviside step function,
${\bf r}_{i}$ (${\bf v}_{i}$) is the location (velocity) of the $i$th vortex,
${\bf r}_{k}^{(p)}$ is the location of the $k$th pinning site, $\xi_{p}$ is
the pinning site radius, $N_{p}$ ($N_{v}$) is the number of pinning sites
(vortices),
${\bf {\hat r}}_{ij} =({\bf r}_{i} - {\bf r}_{j})/|{\bf r}_{i} - {\bf r}_{j}|$,
${\bf {\hat r}}_{ik} =({\bf r}_{i} - {\bf r}_{k}^{(p)})/
|{\bf r}_{i} - {\bf r}_{k}^{(p)}|$, and we take $\eta = 1$.

\begin{figure}
\centerline{
\epsfxsize=3.0in
\epsfbox{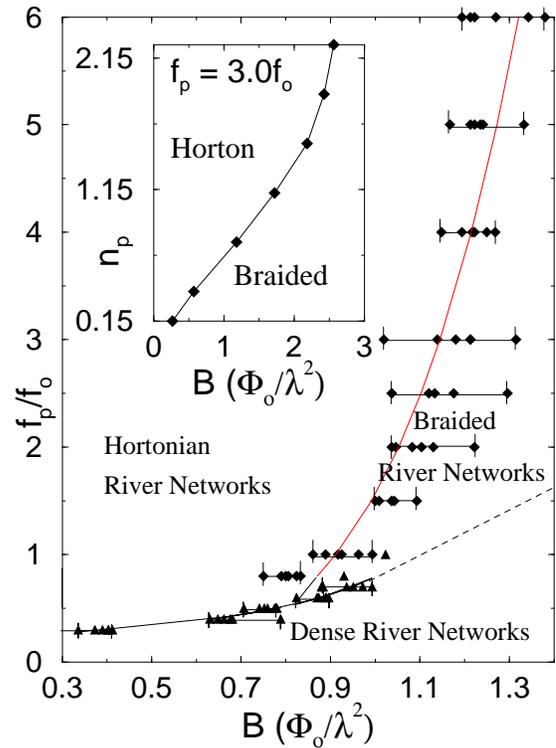}}
\caption{The vortex river network morphological 
phase diagram for pinning force, $f_{p}$, versus
magnetic field, $B$, for $n_{p}=0.75/\lambda^{2}$
(thus, here $B_{\phi} = 0.75 \, \Phi_0 / \lambda^{2}$).
In regions of very low pinning force, dense 
vortex river networks dominate. 
%
%
For higher pinning $f_{p}$'s, the Hortonian 
rivers become braided when $B$ grows.
%
%
For samples with significant amount of pinning, it is
the initial front (with low {\it local\/} density of 
field lines $B$, $B < 3 B_{\phi} / 2$ [7], 
and thus dominant pinning force $f_p$) which branches 
out in a Hortonian manner.  Behind this initial front 
follows the (intermediate-$B$) braided region.  Further 
behind, follows the (large-$B$) dense-flux regime.  
The inset shows the shift in the 
Horton-braided boundary $f_{p}=3.0f_{0}$
as the pinning density, $n_{p}$, is changed.  
As $n_{p}$ is increased, the Horton-braided 
boundary shifts towards higher $B$.
The broad crossover boundaries are in the region of triangles and 
rhombuses.  The (power-law-fit) lines are just guides to the eye.
The dense-braided crossover at high-fields (dashed) 
is an extrapolation of the power-law-fit for low fields;
the former is very difficult to compute because it requires
a large number of vortices monitored over very long times.
}
\label{fig:phase}
\end{figure}

{\it Morphological Characterization.---\/}
In order to identify and characterize the vortex river networks formed
as the flux-gradient-driven front initially penetrates the sample, 
we divide our simulation area into a $300 \times 300$ grid.  
Each time a vortex enters a grid element, the counter associated 
with that element is incremented.  All grid elements that are 
visited at least once by a vortex are considered part of
the network \cite{RRiver}.
The maximum number of vortices in the sample is approximately 1200.
The pinning density $n_{p}$ and radius $\xi_{p}$ were kept constant at
$n_{p} = 0.75/\lambda^{2}$ and $\xi_{p}=0.3\lambda$, while the pinning
force $f_{p}$ varied from sample to sample. 
We also performed additional
simulations in which $f_{p}$ was kept constant and $n_{p}$ varied 
from $n_{p}=0.15/\lambda^{2}$ to $n_{p}=2.15/\lambda^{2}$.
%

We observed three distinct vortex river network morphologies, 
depending on the {\it local\/} magnetic field $B$ 
and the pinning force $f_p$, 
as indicated in one of our main results:
the ``morphological phase diagram" in Fig.~\ref{fig:phase}.
%
In Fig.~1  
the vortex trajectories are presented for the three 
morphologies. In samples with low pinning force values,
$f_{p} \lesssim 0.75f_{0}$ (see Fig.~\ref{fig:phase}),
vortices flow throughout the sample, producing dense
vortex river basins.  These become space-filling 
for large times---or large fields since
the external field is slowly ramped up.
An example of the vortex channels in this regime,
as they appear after 160000 MD steps, is shown in 
Fig.~1(c).
If the simulation is
allowed to proceed for a larger number of MD steps, the
channels eventually fill the entire region shown in 
Fig.~1(c).
For stronger pinning, $f_{p} \gtrsim 1.0 f_{0}$, 
and low vortex densities 
$B \lesssim \, \Phi_{0}/\lambda^{2} \approx 3 B_{\phi}/ 2$,
we observe branched ``Hortonian'' river networks that follow 
Horton's laws of stream number and length 
[see Fig.~1(a)].
At higher magnetic fields
$B \gtrsim \Phi_{0}/\lambda^{2} \approx 3 B_{\phi}/2 $,
the vortex rivers become highly braided or 
interconnected and are no longer Hortonian in morphology 
[see Fig.~1(b)].
Unlike the dense networks of Fig~1(c), where 
preferred vortex paths are uncommon, 
in the braided regime vortices consistently 
move along certain pathways, while in some areas of 
the sample vortex motion rarely occurs.
%
%
%

\begin{figure}
\centerline{
\epsfxsize=3.3in
\epsfbox{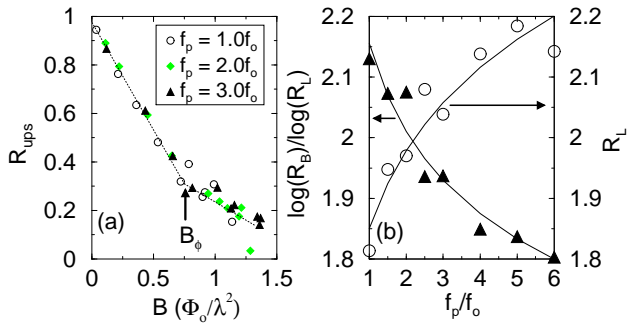}}
\caption{(a) Fraction $R_{\rm ups}$ of unoccupied pinning sites (ups)
versus $B$ for six samples with different $f_p$'s. 
We find a change in the rate at which pins become occupied
with increasing field: it decreases noticeably for $B>B_{\phi}$. 
(b) The length ratio $R_{L}$ and 
$D_{F}/d_{c}=\log{R_{B}}/\log{R_{L}}$, 
versus the pinning force, $f_{p}$.  
The stream dimension, $d_{c}$, is one.
Only the trend in the fractal dimension $D_{F}$ 
can be observed from above because the error bars for 
$R_{L}$ and $D_{F}/d_{c}$ are $\pm 0.1$.  The lines are 
power-law best fit curves, which only provide a guide to the eye.
The formula for $D_F$
gives values slightly above 
2, because it assumes that Horton's laws hold
at all length scales, while our vortex basins
only span a very limited range of length scales.
%
}
\label{fig:trend}
\end{figure}

For low pinning forces in the dense network regime,
vortex motion occurs {\it both\/} interstitially
(with the vortices moving only in the areas between pinning sites)
and by means of depinning.
If vortex depinning is occurring in a landscape with
traps of comparable strength, {\it no\/} favored paths for 
flux motion can form, leading to the observed dense pathways.
The Hortonian and braided regimes arise
once the pinning is strong enough that predominantly
interstitial motion occurs.
That is, pinned vortices almost never depin.
Other vortices are prevented from moving close to a pinned vortex
by the vortex-vortex repulsion, which has a longer 
(by near two orders of magnitude) range than the
attraction of each pinning site.
Since there are regions of the sample
(i.e., at or near pinned vortices) where flux motion does not occur,
the flow of the moving vortices must be concentrated in certain
well-defined regions or rivers, leading to the formation of
either Hortonian or braided rivers.

The broad crossover between Hortonian and braided rivers
occurs when the flux density has increased enough that
a large fraction of the pinning sites are occupied.
In Fig.~\ref{fig:phase}, the crossover region 
increases from $B \approx 0.9\, \Phi_{0}/\lambda^{2}$ for 
$f_{p} = 1.0f_{0}$, to $B \approx 1.3\, \Phi_{0}/\lambda^{2}$ 
for $f_{p} = 6.0f_{0}$.
In each case, the crossover occurs at vortex densities 
higher ($3 B_{\phi} / 2 \lesssim B < 2 B_{\phi}$)
than the matching field 
$B_{\phi}=0.75\, \Phi_{0}/\lambda^{2}$, 
  when $N_p=N_v$ \cite{matching}.
%
This 
is in agreement with the results for the inset of
Fig.~\ref{fig:phase}, which shows that the
transition from Hortonian rivers to braided rivers
occurs at higher vortex densities as $n_{p}$ 
(and thereby the matching field) is increased.  
Additional support for this interpretation comes from
examining the fraction $R_{\rm ups}$ of 
unoccupied pinning sites [Fig.~3(a)].  
At the matching field, $B_{\phi}=0.75 \, \Phi_{0}/\lambda^{2}$,
only about $65\%$ of the pins are occupied [7,11].
The pins are not fully occupied until a field of 
$B \approx 1.4 \, \Phi_{0}/\lambda^{2} \approx 2 \, B_{\phi}$ 
is applied---when 
%
the potential energy landscape experienced by the 
moving vortices becomes much more uniform.

\begin{figure}
\centerline{
\epsfxsize=3.3in
\epsfbox{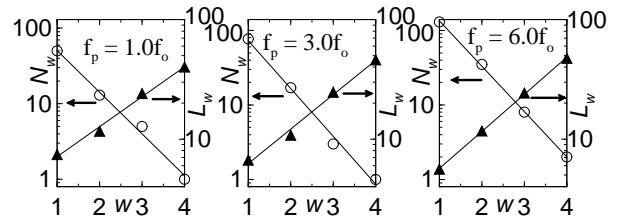}}
\caption{ 
The number of streams $N_{w}$ of order $w$, 
and their lenghts $L_{w}$, for vortex river networks 
with three different pinning forces $f_p$.  {\it Six\/} 
different $f_p$'s gave virtually identical plots---all 
obeying Horton's laws.
}
\label{fig:Horton}
\end{figure}

{\it Horton Analysis.---\/}
In order to determine whether the vortex river networks we observe obey
Horton's laws, we performed Hortonian analysis on five different
realizations of disorder for each of eight different pinning forces
$f_{p}$ falling within the Hortonian regime.  In each trial, a branching
river was identified for analysis.  The numbers $N_{w}$ and lengths $L_{w}$
of streams of order $w = 1$ to 4 were recorded. \cite{limitations}
Representative plots of the
type used to determine the length ratio,
$R_{L}=L_{w+1}/L_{w}$, and the bifurcation ratio,
$R_{B}=N_{w}/N_{w+1}$, are shown in Fig.~\ref{fig:Horton}.
In the best fit exponential regressions used to extract $R_{L}$ and $R_{B}$,
the average correlation coefficient was $~0.99$, indicating a good
fit to the Hortonian relationships.
The average values for $R_{B}$ and $R_{L}$ throughout the Hortonian
river region were $R_{B} = 3.99 \pm 0.18$ and
$R_{L} = 2.04 \pm 0.12$, in excellent agreement with
geophysical rivers. \cite{Rodriguez,randomHorton}.

The characteristics of the Hortonian river networks
are dependent on the pinning force $f_{p}$. In 
Fig.~3(b)
we plot the length ratios $R_{L}$, and fractal
dimensions $D_{F}$, for each pinning force in the Hortonian region.
%
%
The braching ratio (not shown) is roughly constant as $f_{p}$ is varied.
Changing the pinning force alters the ease with which individual
vortices can be depinned, and thereby changes $R_{L}$ and $D_{f}/d_{c}$.
As the pinning force decreases, it is more likely that some vortices
will be depined and form new pathways of vortex motion.
This will decrease the length of the higher order rivers
by cutting short how far the vortex channels propagate before bifurcating.
Therefore $R_{L}$ will decrease with decreasing $f_{p}$.
Since a larger number of paths are created the $D_{f}$ will 
increase with decreasing $f_{p}$, in agreement
with Fig.~\ref{fig:trend}(b).

{\it Concluding remarks.---\/}
We have analyzed the morphologies of flux-flow channels
slowly driven to its marginally stable state, as a function of
flux density and disorder strength. We have identified 
three distinct morphologies\cite{videos} which include: 
a (large $B$) dense network regime, where flow can occur
anywhere; a braided network regime, where flow is restricted 
to certain regions; and a (low $B$) Hortonian network regime, 
where Horton's laws of length and branching ratio 
are obeyed in agreement with geophysical rivers.  
%
%
Indeed, it seems promising to analyze tree-shaped 
channel flow at the microscopic level adapting concepts that 
have already been successful in treating macroscopic river basins.
These types of analysis are largely unexplored.  
The direction and success of such an approach 
constitutes an open and fascinating area.

CJO (APM) acknowledges support from the GSRP of 
the microgravity division of NASA (NSF-REU).
We thank the Maui Supercomputer Center, R.~Riolo, and 
the UM-PSCS
for providing computing resources.  
We thank  
F.~Marchesoni, M.~Bretz, E.~Somfai, D.~Tarboton, 
and S.~Peckham for their comments.


\begin{references}
\bibitem{limitations}
Streams of order $w=5$ and above were very 
rare in our samples because our landscapes are 
flat, with small parabolic potential wells on it.
Thus, the vortex microscopic landscape is very different 
from the quite rough mountain-like landscapes 
used to model some geophysical river basins.  Therefore, 
Horton's laws for vortices do not extend over a wide range 
of length scales.
\bibitem{matching}
This is due to the fact that a 
fraction of pinning sites remains unoccupied at the matching field,
as shown in experiments 
and simulations \cite{RBose}. 
This is a consequence of the random spatial distribution of the pinning.
If one pinning site is very close to a second site containing a vortex,
the repulsive force of the trapped vortex will prevent the empty pin from
being occupied until the vortex density increases further.
\end{references}
\end{document}